# CW HIGH INTENSITY NON-SCALING FFAG PROTON DRIVERS

C. Johnstone, Fermilab, Batavia 60510, USA
M. Berz, K. Makino, Michigan State University, East Lansing, MI USA
P. Snopok, Illinois Insitute of Technology, Chicago, IL USA

*Abstract*

Accelerators are playing increasingly important roles in basic science, technology, and medicine including nuclear power, industrial irradiation, material science, and neutrino production. Proton and light-ion accelerators in particular have many research, energy and medical applications, providing one of the most effective treatments for many types of cancer. Ultra high-intensity and high-energy (GeV) proton drivers are a critical technology for accelerator-driven sub-critical reactors (ADS) and many HEP programs (Muon Collider). These high-intensity GeV-range proton drivers are particularly challenging, encountering duty cycle and space-charge limits in the synchrotron and machine size concerns in the weaker-focusing cyclotrons; a 10-20 MW proton driver is not presently considered technically achievable with conventional re-circulating accelerators. One, as-yet, unexplored re-circulating accelerator, the Fixed-field Alternating Gradient, or FFAG, is an attractive alternative to the cyclotron. Its strong focusing optics are expected to mitigate space charge effects, and a recent innovation in design has coupled stable tunes with isochronous orbits, making the FFAG capable of fixed-frequency, CW acceleration, as in the classical cyclotron. This paper reports on these new advances in FFAG accelerator technology and references advanced modeling tools for fixed-field accelerators developed for and unique to the code COSY INFINITY[1].

## INTRODUCTION

The drive for higher beam power, high duty cycle, and reliable beams at reasonable cost has focused international attention and design effort on fixed field accelerators, notably FFAGs. Ultra high-intensity, high-energy (GeV) proton drivers are a critical technology for applications such as accelerator-driven sub-critical reactors (ADS) and many HEP programs (Muon Collider). High-intensity GeV proton drivers encounter duty cycle and space-charge limits in the synchrotron and machine size concerns in the weaker-focusing cyclotrons. A 10-20 MW proton driver is challenging, if even technically feasible, with conventional accelerators – with the possible exception of a SRF linac, which has a large associated cost and footprint and the ultra-high reliability issues further complicate the accelerator technology with duplicity in the accelerator presently considered the most feasible approach.

One re-circulating candidate, the FFAG is an, as yet unexplored, fixed-field alternative to the cyclotron, and its strong focusing optics should mitigate space charge effects and achieve higher bunch charges than possible in a cyclotron. Recently, the concept of isochronous orbits has been explored and developed for non-scaling FFAGs using powerful new methodologies in accelerator design and simulation. The property of isochronous orbits enables the simplicity of fixed RF and, by tailoring a nonlinear radial field profile, the FFAG can remain isochronous beyond the energy reach of cyclotrons, well into the relativistic regime. With isochronous orbits, the machine proposed here has the high average current advantage and duty cycle of the cyclotron in combination with the strong focusing, smaller losses, and energy variability that are more typical of the synchrotron. With the cyclotron, the current industrial and medical standard, a competing CW FFAG could potentially represent a strong competitor in the broader accelerator community. Further, a low intensity, CW accelerator with variable energy is a desirable innovation in medical accelerators for cancer therapy.

In summary, proton and light-ion accelerators have many research, energy and medical applications, providing one of the most effective treatments for many types of cancer, but the current state of technology in accelerators severely limits their potential use. A high-energy, high-intensity CW re-circulating accelerator would have broad impact on HEP facilities and potentially nuclear power, and open up a range of as-yet unexplored industrial applications. This paper reports on these new advances in FFAG accelerator technology and announces advanced modeling tools for fixed-field accelerators unique to the code COSY INFINITY.

## TYPES OF FFAG ACELERATORS

The FFAG concept in acceleration was invented in the 1950s independently in Japan, Russia and the U.S. The field is weak at the inner radius and strong at the outer radius, thus accommodating all orbits from injection to final energy. Focusing is provided by an alternating body gradient (which alternately focuses in each transverse plane) or through body gradient focusing in one plane (nominally horizontal) and strong gradient-dependent edge focusing in the other (vertical) plane. There are two overarching classifications for FFAG accelerators: the so-called scaling and non-scaling variants

"Work supported by Fermi Research Alliance, LLC under Contract No. DE-AC02-07CH11359 with the US Department of Energy."

Scaling FFAGs: Scaling FFAGs (either spiral or radial-sector FFAGs) are characterized by geometrically similar orbits of increasing radius imposed by applying magnetic fields that follow a scaling law as a function of radius: $B(r, z) = B_0(z) (r/r_0)^k$, where k is the FFAG field index. Direct application of high-order magnetic fields and edge focusing maintains a constant tune and other constant optical functions (such as zero chromaticity) during the acceleration cycle, thus avoiding low-order resonances. Recently Y. Mori (Kyoto University Research Reactor Institute, Japan) has revitalized scaling FFAGs, building and operating several technically modern versions[2]. His work has been applied to commercial FFAGs under development at both Mitsubishi Corp. and NHV Corp.

Non-scaling FFAGs: The scaling condition is relaxed in the non-scaling FFAG with stable acceleration the primary goal. Initially non-scaling designs had a large tune dependence with momentum, which limited beam lifetimes to on the order of ten turns [3]. The first non-scaling FFAG - the Electron Model for Many Applications (EMMA)[4] has been built at Daresbury Laboratory - and has attracted international attention by their recent success[5] in serpentine, or bucketless, acceleration from ~15 to 18 MeV through resonances as shown in Figure 1 (the twisted phase space is characteristic of this very specific type of non-scaling FFAG). Subsequent non-scaling designs were developed with constant or acceptably small-variations in tunes. These recent innovations in non-scaling FFAG design sustain the slow, multi-turn stability needed for a proton driver. [6,7]. Specific tune-stable and isochronous non-scaling FFAG designs are the subject of this work.

constant tune and is not suitable for an accelerator with a modest RF system and therefore a slower acceleration cycle - its application is rapid acceleration (of unstable particles such as muons, for example).

Since the reference orbit in a fixed-field accelerator moves with energy, tune can be controlled using a linear or nonlinear gradient FFAG and by shaping the edges of the magnets. All three focusing terms (gradient, edge and centripetal or weak focusing) are impacted by the edge contour and their interaction can be used to manipulate the machine tune in the horizontal. Two terms, gradient and edge focusing, are available for tune control in the vertical. Further, in a non-scaling FFAG, contributions from the different strength terms can vary with radial position and can also be independent in the F and D magnets. This independence plus an increase in strength of all the terms with radius tracks the increase in momentum and stabilizes the tune. The result is a dramatic enhancement in the momentum reach of the machine, from 2-3 to a factor of 6 utilizing a simple edge contour and ultimately to a factor of 44 by exploiting additionally a nonlinear gradient. An arbitrary field expansion has been exceptionally successful in controlling both tunes and physical attributes of a machine and with surprisingly large dynamic acceptance. Isochronous orbits were achieved in a non-scaling FFAG by applying both a nonlinear gradient and edge contour, thereby achieving CW operation and simple RF systems.

The integrated B field must be a nonlinear function of radius to keep it proportional to the relativistic velocity - the condition for isochronous orbits, and, again, with the tune adjusted via an edge contour. Unlike the cyclotron which relies on a dipole field and is therefore limited in adapting path length to match relativistic velocities, the non-scaling FFAG can maintain isochronous orbits at strongly relativistic energies as shown in Figure 2.

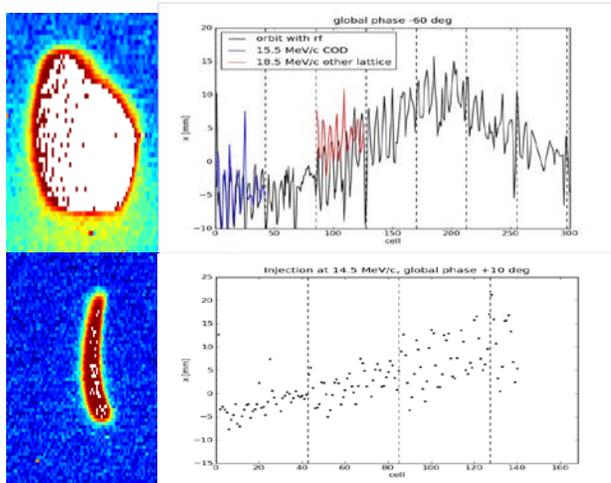

Figure 1. No acceleration (left, top) vs. accelerated beam (right, bottom) in EMMA, a linear-field, rectangular magnet-edge machine. Also shown are measured orbit displacements at fixed energy and measured orbit movement during acceleration[5].

## Progress in Non-scaling FFAG Design

Initial non-scaling FFAG lattices (EMMA project)[4] utilized a linear fields/constant gradient and rectangular magnets. This implementation does not maintain a

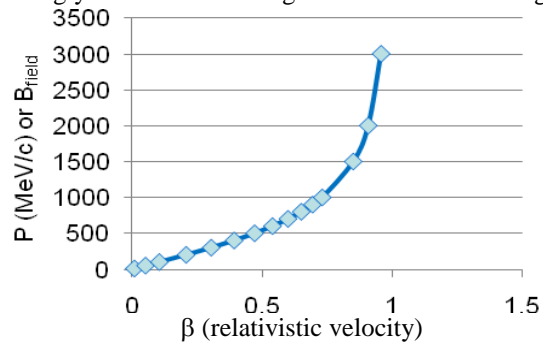

Figure 2. Momentum dependence ($\propto$ <B> field) on velocity (or path length) required to maintain the isochronous condition.

## Isochronous FFAG Accelerator

As discussed above, the concept of isochronous orbits has been invented for non-scaling FFAGs. This concept has been tested on a preliminary 0.25-1 GeV non-scaling FFAG designed using new methodologies and optimizers[8] (Figure 3-4 and Table 1). A complete CW accelerator system would likely entail an H⁻ injector. (Use of H- in the lower energy ring permits CW injection into the higher-energy ring through charge-changing or stripping methods.)

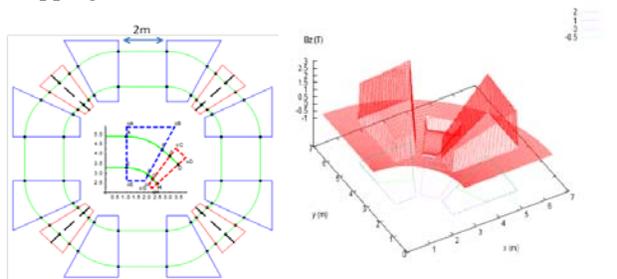

Figure 3: Ring layout and 3D field profile from COSY. The 3D field profile of a quarter of the ring generated by the new tools in COSY INFINITY expanded from a simple hard-edge radial field profile to a full radial/azimuthal distribution with fringe fields

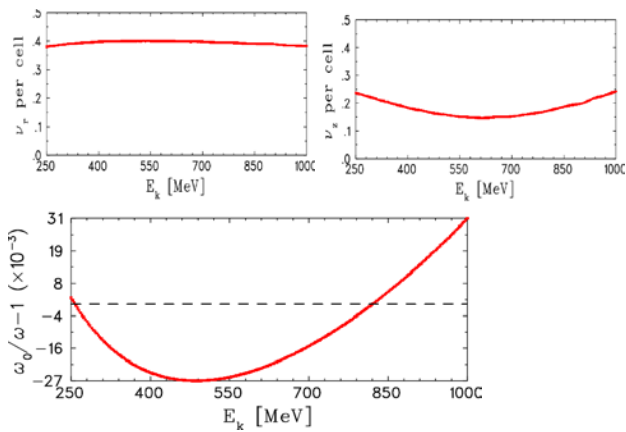

Figure 4. Results using the cyclotron code CYCLOPs showing radial tune per cell (top, left), azimuthal tune per cell (top, right), and frequency change in percent (bottom)[9].

Table 1: Parameters of a 0.25-1 GeV isochronous FFAG.

| Parameter | Value |
| --- | --- |
| Circumference | 32m, 4 cells |
| Cell Tunes ($\nu_x, \nu_y$,) | 0.38, 0.24 |
| F,D lengths (inj-ext) | 1.17-1.94, 0.38-1.14 m |
| Field B, F,D(inj-ext) | 1.6 - 2.4T, –0.1 - –0.42T |
| Magnet Aperture (H,V) | 1.6m, 5 cm |

## FFAG DESIGN AND MODELING

Powerful new methodologies in accelerator design and simulation have been pioneered using control theory and optimizers in advanced design scripts with final simulation in COSY INFINITY which now has a full complement of sophisticated simulation tools to fully and accurately describe both conventional accelerators and the FFAG's complex electromagnetic fields[8].

The preliminary results (Figure 5) of these initial studies indicate stable tunes and large dynamic apertures – additional optimization will establish final desired machine parameters, and, more importantly, the results indicate a strong degree of isochronous operation. This lattice proves to be a viable starting point for development of an isochronous FFAG with either a fixed, or a rapidly modulated, RF system.

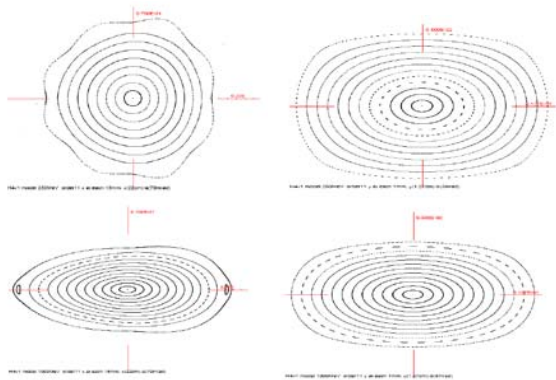

Figure 5: Dynamic aperture at 0.25, and 1 GeV – step is 15 mm in the horiz. (left) and 1 mm in the vert. (right).